# Recoilless Resonant Capture of Antineutrinos

R. S. Raghavan

*Institute for Particle Physics & Astrophysics and Department of Physics*
*Virginia Polytechnic Institute and State University, Blacksburg VA 24061*

Resonant capture of antineutrinos ($\tilde{\nu}_e$) can be accomplished by exploiting the monoenergetic $\tilde{\nu}_e$ emitted in bound state β-decay. Extending this idea, I explore conditions for recoilless resonant capture in the system $^3$H - $^3$He. Observation of such transitions can set the stage for placing stringent limits on the $\tilde{\nu}_e$ parameter $\theta_{13}$ on an ultra-short baseline of ~9 m and for observing the gravitational red shift of neutrinos.

The antineutrino ($\tilde{\nu}_e$) capture reaction

$$\tilde{\nu}_e + A(Z) + e^-(\text{orbital}) \rightarrow A(Z-1) \qquad (1),$$

where the electron is captured from atomic orbit in the target atom A(Z) was considered by Fermi[1] and by Mikaelyan et al [2] who noted the resonant character of the reaction. The resonance energy is:

$$E(\tilde{\nu}_{e\,res}) = M(Z) - M(Z-1) = Q \qquad (2),$$

where M are the *atomic* masses of the target and daughter atoms. The rate R of (1) is[2]:

$$R \propto |\Psi^2| \, \rho(E\,\tilde{\nu}_{e\,res}) / ft \qquad (3),$$

where $|\Psi^2|$ is the probability of the orbital electron in the nucleus (as in normal orbital electron capture decay), $\rho(E\,\tilde{\nu}_{e\,res})$ is the resonant spectral density of the inducing $\tilde{\nu}_e$ beam (the number of $\tilde{\nu}_e$ per unit energy interval at the resonance energy) and ft is the reduced half-life of the β-decay A(Z-1)→A(Z). Ref. 2 considered an incident $\tilde{\nu}_e$ spectrum from nuclear reactors that falls monotonically up to 10 MeV. Reaction 1) has never been attempted primarily because $\tilde{\nu}_e$ beams from normal radioactive sources or reactors (the only known $\tilde{\nu}_e$ sources), have too small a resonant spectral density $\rho(E\,\tilde{\nu}_{e\,res})$.

In this paper I propose a new approach that could make reaction (1) observable (perhaps even in the laboratory) by significantly increasing $\rho(E\,\tilde{\nu}_{e\,res})$. The key idea is a way to achieve a source of *monoenergetic* $\tilde{\nu}_e$'s with the precise resonance energy required in (2). Such a source is the reverse decay:

$$A(Z-1) \leftarrow \tilde{\nu}_e + A(Z) + (e^- \text{ in bound state}) \qquad (4)$$

in which the electron emitted in the usual β-decay of A(Z-1) occupies a bound orbit in A(Z) instead of going into the continuum. This process, called bound state β-decay (Bβ), was recognized already in 1947[3] as a competing mode to the normal continuum mode (Cβ). The definitive work on Bβ in the modern V-A theory is due to Bahcall[4] whose results apply directly to the present work. If the electron final state in the $\tilde{\nu}_e$ source (4) is the same atomic orbital as that from which it is captured in the target of (1), the resulting $\tilde{\nu}_e$ is monoenergetic due to the 2-body decay (4) and its energy = M(Z)-M(Z-1)=Q, precisely the resonance energy (2). The basic condition for truly resonant $\tilde{\nu}_e$ capture can thus be satisfied.

The purpose of this Letter is to explore $\tilde{\nu}_e$ resonance reactions in the favorable case of the $^3$H (T)-$^3$He system. We extend these studies to explore basic parameters and experimental conditions for dramatically enhancing the resonance cross sections via *recoilless* $\tilde{\nu}_e$ emission and absorption. Possible immediate applications are the study of $\tilde{\nu}_e$ phenomena such as $\theta_{13}$ oscillations on a very short (laboratory scale) baseline and the gravitational red shift of neutrinos.

The nuclear data on the T-$^3$He system is summarized in Table 1. This case satisfies all the basic requirements for observing resonant

Table I. The T→³He nuclear-atomic system

| Decay | E$\tilde{\nu}_{eres}$ (keV) | ft(s) (Cβ) | Bβ/Cβ (ref. 4) |
|---|---|---|---|
| ³H(T)→ ³He | 18.60 | 1132 | 4.7 x10⁻³ |

$\tilde{\nu}_e$ capture. It offers the fastest known super-allowed β-decay that emits one of the lowest energy $\tilde{\nu}_e$'s with a sizable Bβ/Cβ ratio. Most importantly, the atomic physics basis is ideal. The initial T atom has a vacancy in the 1S shell for Bβ decay and the target ³He has two 1S electrons one of which can be captured. 1S electrons are ideal since $|\Psi|^2$ (which appears both in the Bβ and the $\tilde{\nu}_e$ capture probabilities), is maximized with the largest overlap of the radial wave function with the nucleus and offers n =1 for the atomic principal quantum number ($|\Psi|^2 \propto 1/n^3$).

The resonant $\tilde{\nu}_e$ capture cross section is[2]:

$$\sigma = 4.18 \times 10^{-41} g_o^2 \rho(E \tilde{\nu}_{e\ res}/\text{MeV}) / ft \ \text{cm}^2 \quad (5)$$

with $g_o^2 \sim 4(\alpha Z)^3 \sim 1.2 \times 10^{-5}$ for 1S electrons. What is the spectral density ρ? Consider a source of T and a ³He target. at 300K. In this case, the $\tilde{\nu}_e$ microspectrum will have a Doppler profile due to the motion of the T atoms centered around the resonance energy of 18.6 keV with a FWHM $2\Delta = 2E_\nu(2kT/Mc^2)^{1/2} = 2 \times 18.6 \times 10^3 (2 \times 300 \times 8.6 \times 10^{-5}/3 \times 10^9) = 0.11$ eV which applies to source and target. Thus ρ in equation (5) is ~10⁶/0.11. The incident $\tilde{\nu}_e$ spectrum overlaps only partially with the resonance window in the target because of the nuclear recoil with energy $E_R = E_\nu/(2Mc^2) = 0.055$ eV. The resonance cross section is thus[2]:

$$\sigma(res) \sim 4.18 \times 10^{-41} \times 1.2 \times 10^{-5} \times 9.1 \times 10^6 \times 0.25 / 1132 \sim 10^{42} \text{ cm}^2 \quad (6)$$

where the spectral density is given by the Doppler width 2Δ=0.11 eV, the factor 0.25 is the resonance overlap of the Doppler $\tilde{\nu}_e$ profiles at source/target, each of which is recoil shifted by $E_R$ (~Δ) away from the resonance energy. Note that the cross section of the normal ($\tilde{\nu}_e + p \rightarrow n + e^+$) reaction is ~9x10⁻⁴³ cm² at $E_\nu$ = 3 MeV. Comparing the two reactions, three basic points emerge: 1) The resonance cross section (6) is as large as σ($\tilde{\nu}_e + p$) even though the $\tilde{\nu}_e$ energy is ~150 times smaller. 2) The proposed idea shows how to observe for the first time, $\tilde{\nu}_e$ reactions well below the $\tilde{\nu}_e + p$ threshold of 1.8 MeV. 3) The immediate consequence is the attractive possibility of ultra-short baselines for measuring the $\tilde{\nu}_e$ parameter $\theta_{13}$. The optimum baseline for this measurement, L(m)/E(keV) = 0.5, is only 9.3 m for the T(Bβ)+³He resonance reaction in contrast to 1500 m for $\tilde{\nu}_e + p$. Clearly, such a prospect encourages deeper study of the T-³He resonance reaction.

As a benchmark for a (TBβ-³He) $\theta_{13}$ measurement on a 9.3 m baseline we estimate using the data of Table 1 and (6), a capture rate: R = 3x10⁻²/day for 100MCi T source and a 1kg ³He target. The modest rate is due to the modest (for a $\tilde{\nu}_e$ experiment) target mass. For initial experiments at close geometries at ~10 cm the rate is ~10⁴ times higher.

Two methods are available for measuring the reaction signal. First, the β-activity of the accumulated T can measured if the capture rate is high. For the close geometry case (10 cm), the β rate is S = 3x10²/ (τ(T)= 7000d) =0.04/d *initially* but increasing linearly with time. The time dependence of the activity bestows a useful signature for $\tilde{\nu}_e$ capture. Secondly, the reaction rate could be measured by determining the number of T atoms accumulated in a given time. The classic reasons for accelerator mass-spectroscopy of long lived radioactivity (such as ¹⁴C) directly apply in the present case. Separation of T from ³He may be easier than ¹⁴C dating because of the x2 charge difference of T and He ions. The main background in either case will be (n,T) reactions thus requiring heavy neutron shielding. However, background from all sources can be directly measured in a source-out procedure.

The question now arises: Is it possible to enhance the spectral density $\rho(E \tilde{\nu}_{e\ res})$ even further to increase the cross-section (6)? The question leads logically to *recoilless $\tilde{\nu}_e$ emission and absorption* in the TBβ→³He system. In that case: a) the absence of recoil eliminates the resonance mismatch in (6) and b) with proper design, the $\tilde{\nu}_e$ microspectrum could have a line width $<< 2\Delta$. On the other hand, the probability of recoilless transitions critically depends on incorporating the source and target atoms in favorable *solid* lattices. The crucial recoilless transition probability is perforce very low because the low-mass T and $^3$He suffer relatively energetic recoils that must be suppressed by crystal forces. The discussion below illustrates the interplay of parameters.

The probability of the recoilless effect is given by $f = f_1(T)f_2(He)$. The recoilless fraction $f_1$ or $f_2$ is given by[5]

$f_x = \exp{-[3E_R/2k\Theta_3\{1+ 0.25(T/\Theta_3)^2 \mathcal{D}(T/\Theta_3)\}]}$. (7)

$\Theta_3$ is the effective Debye temperature of the mass 3 atom in a lattice of mass M atoms,. T is the ambient temperature and $\mathcal{D} = (\int x dx /(e^x - 1))$ taken from 0 to $x = T/\Theta_3$).

Estimates of $\Theta_3$ and thus f depend on the nature of the T and $^3$He when incorporated in solids. A huge literature is available on loading $^1$H, T, $^3$He and He in metallic lattices..[6] Tritium forms tritides in metals. Thus one could estimate $\Theta_3$ guided by Mőssbauer data of light and heavy impurities such as Fe[7] and Sn[8] in various host lattices. The results show that the effective $\Theta$ roughly follows the rule[9] $\Theta_{eff} \sim \Theta_{host}$ x $(M/M')^{1/2}$ where M is the host mass and M' the impurity mass. For M'=3, this rule would indicate a substantially higher $\Theta_3$ for most hosts. Lacking detailed specific information, I shall adopt for safety, simply $\Theta_3 \sim \Theta_{host}$ in Table 2 for two host examples Al and Be. Helium loading is a different situation since it has no solubility in solids and chemical methods fail[10]. In this case, ion implantation offers a way out. Helium has been implanted in common metals to several percent and it has remained stable for years at or below 300K.[11] The nature of the He in metals is very different from T. It forms ~10nm microbubbles with gas pressures exceeding 1 GPa[12]. We note that recoilless *γ-rays* have been observed for noble gases such as Kr[13] and Xe[14] trapped in microcages in clathrates which may be similar to microbubbles in metals. We arbitrarily assign a low $\Theta_3$ ~100K for He The recoilless fractions fraction $f = f_1f_2$. calculated for the assumed $\Theta_3$ are given in Table 2. We stress again that the assumed values above are crude.

Table 2. Probability $f = f_1f_2$ for recoilless transitions for mass 3 atom in host lattices.

|    | $\Theta_3$ (K) | T (K) | $f_1$ | $f_2$ | f |
|----|------|-----|------|---------|-----------|
| Be | 100  | 50  |      | ~$10^{-4}$ |           |
| Al | 143  | 80  | 0.11 |         | ~$10^{-5}$ |
| Be | 520  | 300 | 0.35 |         | ~$3.5 \times 10^{-5}$ |

The next vital parameter is the line width in recoilless emission and absorption. The relevant width is the practical energy width for H or T in solid lattices determined by energy fluctuations via interactions of the nuclear moments with ambient fields. Fortunately, the quadrupole moments of $^3$H and T are zero. Magnetic interactions are minimized in diamagnetic or weakly paramagnetic lattices. Magnetic relaxation times $T_1$ have been measured by NMR for (Pd-H$_{0.7}$) [15] and for $^3$He microbubbles accumulated in aged T(Pd) samples[12]. These values of $T_1$ are directly usable as recoilless line widths. Typical measured values of $T_1$ at 300K are ~10 ms for H(Pd) and ~140 ms for $^3$He(Pd). A 10ms line width is an energy width of ~$6.6 \times 10^{-14}$ eV, compared to $2\Delta$ ~0.11 eV in free recoil.

If these measured relaxation times are typical, then it is clear that the spectral density is enhanced hugely relative to free recoil, by a factor ~$1.7 \times 10^{12}$. Thus even for $f \sim 10^{-5}$ as in Table 2, the recoilless resonant reaction rate is $1.7 \times 10^7$ times larger than the free-recoil resonant reaction rate. This allows more favorable experimental compromises e.g., reducing the source activity to ~10 MCi and the target mass, say to 1g. Then, the capture rate for recoilless $\tilde{\nu}_e$'s is 50/d (at 9.3m). and ~$5 \times 10^5$ /day (at ~10 cm). With these rates, the initial reverse β-decay signal is S ~0.07./d (9.3m) and 70/d (10cm) that approach practicality (see Table 3 for a summary of all rates).

These prospects open new vistas on neutrino experiments. Besides ultrashort baseline $\theta_{13}$

Table 3. Capture and reverse β signal rates for resonant free-recoil & recoilless $\nu_e$ capture

| | Width (eV) | f | Cap. Rate /d 9300 cm | Cap. Rate/d 10 cm | Init. Rev. β /d 10 cm | T (MCi) | $^3$He (g) |
|---|---|---|---|---|---|---|---|
| Free Recoil | 0.11 | 1 | $3.5 \times 10^{-2}$ | $3.5 \times 10^2$ | 0.05 | 100 | $10^3$ |
| Recoil-less | $6.6 \times 10^{-14}$ | $10^{-5}$ | 50 | $5 \times 10^5$ | 70 | 10 | 1 |

measurements, another exciting application can be foreseen. The extreme precision of the energy of the recoilless TBβ $\tilde{\nu}_e$ implies a $Q = E/\Delta E = \sim 3 \times 10^{17}$, some five orders of magnitude sharper than the famous Mőssbauer resonance in $^{57}$Fe. The application to measuring the gravitational red shift of neutrinos is then immediately evident. The available precision of $\Delta E_\nu/E = 3 \times 10^{-18}$ compared to the neutrino red shift $(\Delta E_\nu)_g/E_\nu = 10^{-18}$/cm fall, allows the measurement of $(\Delta E_\nu)_g$ in a bench top set-up with a fall baseline of a few cm (instead of ~2200 cm in the classic Pound-Rebka experiment). The real-time reverse β signal rate of ~7000/d in close geometries after a 100 day exposure may allow scanning of the resonance with Doppler velocities ~ nm/s. Such low speeds may require compromises on close geometries and the S rate above.

In summary, we have pointed out in this paper the application of $\tilde{\nu}_e$'s from Bβ-decay as a path breaking way to observe very low energy $\tilde{\nu}_e$ capture in a resonant mode. We described the highly favorable basis of the T-$^3$He system for accomplishing this task. We showed that the capture rates could be enhanced dramatically by recoilless transitions if the source and absorber are embedded in metals in which the T and $^3$He relaxation times $T_1$ are of the order of ~10 ms. Ion implanted T and $^3$He in beryllium may offer a promising method for realizing these ideas. Many of the solid state estimates assumed here are crude and must be tested in practice in the chosen source and absorber by NMR and neutron scattering. The current availability of data and expertise in these questions bodes well for rapid technical development for implementing our ideas. The successful observation of recoilless $\tilde{\nu}_e$ resonance capture (initially in high rate close geometries) would demonstrate the process and open the way to important applications.

One in particular is attempting stringent limits on $\theta_{13}$ in a laboratory scale experiment with baselines of ~9 m instead of 1500 m as needed at present for $\tilde{\nu}_e + p$ experiments with reactor $\tilde{\nu}_e$'s. It would also open the dramatic possibility of observing the effect of gravity on neutrinos. Hopefully, recoilless neutrinos would be as fruitful as recoilless γ-rays.

I dedicate this paper to the memory of John N. Bahcall.